\documentclass[journal, 11pt]{IEEEtran}
\usepackage[utf8]{inputenc}
\usepackage[T1]{fontenc}

\usepackage{graphicx}
\usepackage{float}

\usepackage[super]{nth}
\usepackage[normalem]{ulem}
\usepackage[noadjust]{cite}
\usepackage[hidelinks]{hyperref}
\usepackage[shortlabels, inline]{enumitem}

\usepackage{array, makecell, multirow, booktabs}
\setcellgapes{3pt}

\usepackage{amsmath, amsthm, amsfonts, mathtools, optidef}

\usepackage[linesnumbered, algoruled, boxed, lined]{algorithm2e}
\let\oldnl\nl
\newcommand{\nonl}{\renewcommand{\nl}{\let\nl\oldnl}}

\newcommand{\eg}{\textit{e.g.,~}}
\newcommand{\ie}{\textit{i.e.,~}}

\begin{document}

\title{A Network-based Compute Reuse Architecture for IoT Applications}

\author{
	\IEEEauthorblockN{
		Boubakr Nour, \IEEEmembership{Member, IEEE}, and
		Soumaya Cherkaoui, \IEEEmembership{Senior Member, IEEE}
	}
	
	\thanks{
		B. Nour and S. Cherkaoui are with are with the INTERLAB Research Laboratory, Faculty of Engineering, Department of Electrical and Computer Science Engineering, Université de Sherbrooke, Sherbrooke (QC) J1K 2R1, Canada 
		(e-mails: boubakr.nour@usherbrooke.ca, soumaya.cherkaoui@usherbrooke.ca).}
}

\maketitle

\begin{abstract}
	The tremendous advancements in the Internet of Things (IoT) increasingly involve computationally intensive services. These services often require more computation resources than can entirely be satisfied on local IoT devices. Cloud computing is traditionally used to provide unlimited computation resources at distant servers. However, such remote computation may not address the short-delay constraints that many of today’s IoT applications require. Edge computing allows offloading computing close to end users to overcome computation and delay issues.  Nonetheless, the edge servers may suffer from computing inefficiencies. Indeed, some IoT applications are invoked multiple times by multiple devices. These invocations are often used with the same or similar input data, which leads to the same computational output (results). Still, the edge server willfully executes all received redundant tasks. In this work, we investigate the use of the computation reuse concept at the edge server. We design a network-based computation reuse architecture for IoT applications. The architecture stores previously executed results and reuses them to satisfy newly arrived similar tasks instead of performing computation from scratch. By doing so, we eliminate redundant computation, optimize resource utilization, and decrease task completion time. We implemented the architecture and evaluated its performance both at the networking and application levels. From the networking perspective, we reach up to an 80\% reduction in task completion time and up to 60\% reduction in resource utilization. From the application perspective, we achieve up to 90\% computation correctness and accuracy.
\end{abstract}


\IEEEpeerreviewmaketitle

\section{Introduction}
\label{sec:introduction}
The Internet of Things (IoT) is now pervading our daily life with different devices and applications. IoT devices are capable of providing various features (\ie sensing and data collecting), yet are constrained in memory and processing capabilities. Meanwhile, IoT applications become smarter at providing personal assistance and autonomous decision-making by interacting with the environment and collecting and making sense of data ~\cite{samuel2019making}.
Smart cities, for example, are focused on developing, deploying and promoting sustainable development practices. This is largely achieved through compute-intensive and time-sensitive applications running on top of reliable and fast communication technology, such as 5G wireless networks. 

Indeed, many IoT applications  require a very short delay and response time, with high computing resources for execution; which require a join optimization from the network level towards the service endpoint~\cite{kunze-coinrg-transport-issues-04}. Computation and delay related constraints cannot be guaranteed with limited IoT computation capabilities. This gave rise to many paradigms, such as in-network computing~\cite{mai2020network} and edge computing~\cite{filali2020multi}. In contrast with cloud computing where computing servers reside away from end-users, edge computing offers computation near end-users, which minimizes the response delay~\cite{song2020multiobjective}. Although edge computing can satisfy the low-latency constraints in various domain-specific applications (\eg healthcare services, telecommunications, industry 4.0, etc.), it still has limited resources compared to cloud computing and cannot host all cloud's services~\cite{ahmed2017bringing}.

Users, through repeated invocations of applications, engage in the smart city ecosystems in a variety of ways using smartphones, mobile devices, connected cars and connected homes~\cite{rodulfo2020smart}. For instance, connected cameras in:
\begin{enumerate*}[(i)]
	\item smart home perimeters can detect and recognize objects, people, and threats;
	\item smart vehicles can detect pedestrians and bicycles, retrieve information from road signs;
	\item smart streets can detect threats and prevent crime;
\end{enumerate*}
in either scenario, the system will take actions accordingly, which in return contributes to public safety monitoring, protection, and emergency support. These applications are widely used by the public and private sectors and involve heavily repeated invocations.

Meanwhile, with a deep look at the applications' context and semantic, high redundancy of computations can be found. The same application/service is invoked multiple times by multiple users using similar, if not the same, input data~\cite{lee2019case}. For instance, home cameras detect the same people (\eg mostly house residents, neighbors, couriers, etc.) passing by or crossing-over multiple times. Vehicle cameras detect the same road signs multiple times along the drive journey, and the same signs are detected by multiple vehicles on the same driving lane. 

It can be seen that these applications are heavy in terms of computation and are invoked multiple times using redundant inputs. Eliminating computation redundancy will help in further minimizing the delay and optimizing the resource utilization. This can be done, regardless of the application's processing logic, from the networking perspective, \eg at the edge server~\cite{nour2020computeless}. 
To address this matter, we complement edge computing with computation reuse. Computation reuse is no more than computation caching~\cite{sanadhya2012asymmetric}.

Unlike existing works that focus on exact matches between the task's input-output data at the application level~\cite{kannan2016seesaw, abouaomar2018users}, we focus in this work on non-identical input data at the networking level. The inputs, in our study case, are correlated either contextually, spatially, or temporarily, and mapped to the same output. For example, two smart vehicles on the same lane detect the same pedestrian using their cameras, yet the two captured snaps are not similar but yield the same output. 

Computation reuse is built on storing the previously executed tasks at edge servers, and re-using these outputs to execute the newly incoming similar tasks. Here, we differentiate three main research questions:

\begin{itemize}
	\item how to integrate the computation reuse at the networking, rather than the application level?
	\item how to identify if two task inputs are similar and lead to the same output?
	\item how to ensure the correctness of computation reuse?
\end{itemize}

\noindent To address these issues, we:
\begin{itemize}
	\item \textit{extend} the edge forwarding to provide a network-based compute reuse forwarding scheme. Forwarding operations at the core routers are more complex and costly since routers have been designed to forward and switch packets instead of providing additional computation. However, edge servers can perform such operations without significant performance degradation.
	\item \textit{apply} a locality-sensitive hashing function to identify similar inputs. Conventional hashing algorithms hash similar inputs to different values tending to avoid the collision. Locality-sensitive hashing, in particular, tends to generate similar hash values for similar input data.
	\item \textit{study} the correctness of the computation reuse output. Since the input data might be similar and not identical, the output results could not be intact with the computation logic. The correctness of output also defines the achieved gain of using the computation reuse and gauges its efficiency and scalability.
\end{itemize}

Specifically, we design a new architecture that adopts computation reuse at the network level, so that we minimize not only the computation at the edge but also the communication since we eliminate the transmission of large input data to distant cloud servers. Indeed, we extract features from the application's raw data, and then apply a locality-sensitive hashing function that aims at hashing similar inputs to the same hash value. Then, we use hash-tables for a fast lookup process to find similar output to satisfy the newly received tasks.

The rest of the paper is organized as follows.
Section~\ref{sec:design} presents the system model and details the proposed architecture and its working principle.
Section~\ref{sec:evaluation} discusses the evaluation performance and obtained results.
In Section~\ref{sec:related_work}, we review some of the existing works and efforts.
Finally, we conclude the paper in section~\ref{sec:conclusion}.

\section{Network-based Compute Reuse Architecture}
\label{sec:design}
In this section, we present the design of a network-based compute reuse architecture, a practical architecture built on top of edge computing assisted with in-network computing. The proposed architecture features a computation reuse-aware design that aims to reduce the utilization of computing resources and the completion/execution time of tasks.

\subsection{Objective \& Design Goals}
In-network computation is not a new aspect in computer networking. Different attempts have been presented to provide caching and computation at the network level, such as Content delivery network~\cite{chen2019computing} and Named Function Networking~\cite{mtibaa2018towards}. Nevertheless, the logic behind the computing is at the application level. The motivation behind the network-based computation reuse presented here is not only to provide computation reuse at the edge network but also to:
\begin{enumerate*}[(i)]
	\item design a unified and distributed architecture that fits with different IoT applications and use-cases regardless of their computing logic;
	
	\item exploit the computing resources at the edge with lower communication cost;
	
	\item enable partial computation at the network via task decomposition; and
	
	\item support dynamic resource allocation and continuous adaptation to tackle the service needs.
\end{enumerate*}

Thus, the primary goals of network-based computation reuse are to:
\begin{enumerate*}[(i)]
	\item minimize the task's completion time by increasing the possibility of reusing previously executed tasks;
	
	\item minimize the communication cost by reducing the amount of content traversing the core network; and
	
	\item optimize the computation utilization and load.
\end{enumerate*}

\subsection{System Model \& Assumptions}
In this work, we consider a service provider that controls a distant cloud server $c$ and a set of geographically distributed edge servers $E$ . Table~\ref{tab:notation} summaries the most used notations in the paper.

The cloud server has  immense computation resources to execute heavy tasks invoking a specific service. To meet latency constraints and satisfy end-users' quality of experience, the service provider offloads its services to an appropriate edge server $e \in E$, as close to end-user possible. The edge server has enough resources to execute the received tasks, yet has limited resources compared to the cloud capacity. It is important to highlight that in this work we are not designing an offloading scheme. Instead, we center our investigation on computation reuse at the edge itself.

\begin{table}[!b]
	\centering
	\makegapedcells
	\caption{Summary of notations.}
	\label{tab:notation}
	\begin{tabular}{l l}
		\toprule
		\textit{Notation} &
		\textit{Description} \\
		\midrule
		
		$T^s$ & List of tasks invoking service $s$ \\
		
		$f^c$ & Computation capacity on the cloud \\
		$f^e$ & Computation capacity of an edge slice \\
		$b_c$ & Bandwidth between user and cloud \\
		$b_e$ & Bandwidth between user and edge \\
		
		$I_t$ & Task input data \\
		$F_t$ & Task complexity \\
		$O_t$ & Task output data \\
		
		$\Gamma(t)$ & Communication cost \\
		$\chi(t)$ & Computation cost \\
		$\eta(t)$ & Reuse cost \\
		$\zeta_t$ & Completion cost \\
		$L$  & Lookup cost \\
		
		$x_t^e$ & 1-0 Offloading parameter:\\
		& $x_t^e = 1$: task computation is offloaded to edge $e$ \\
		
		$r_t$ & 1-0 Reuse parameter:\\
		& $r_t = 1$: full competition is performed \\
		
		$\gamma_t$ & 1-0 Computation reuse variable:\\
		& $\gamma_t = 1$: computation reuse is performed \\
		\bottomrule
	\end{tabular}
\end{table}

We assume that the edge server receives tasks for execution following the Little's Law $M/M/1$ queuing system. Let $T^s$ denote the list of tasks invoking the service $s \in S$. A task $t$ is defined with the following tuple:
$$ t = \langle~I_t, F_t, O_t~\rangle,$$

\noindent where 
$I_t$ is the task's input data to execute the target service, 
$F_t$ is the execution complexity and defines the required resources to execute the given task with the provided input data, and
$O_t$ is the output data (result) after successfully executing the service with the given input.

In this work, we assume that end-users have limited resources to execute tasks locally. Thus, tasks will be executed either at the edge level, at the edge with computation reuse, or at the cloud level. Having said that, we define three costs: the task's communication cost $\Gamma(t)$, the task's execution cost $\chi(t)$, and the task's reuse cost $\eta(t)$.

\vspace{0.2cm}
\textbf{Communication Cost.}
The task's communication cost is based on the size of task's input and output data ($I_t,~O_t$), and the minimum bandwidth between end-user and cloud ($b^c$), or edge server ($b^e$), respectively. The communication cost is defined as shown in Eq.~(\ref{eq:communication_cost}).

\begin{equation}
	\label{eq:communication_cost}
	\Gamma(t) = x_t^e.\frac{I_t + O_t}{b^e} + (1-x_t^e).\frac{I_t + O_t}{b^c}
\end{equation}

\noindent where $x_t^e$ denotes whether a task $t$ is executed at the edge server ($x_t^e = 1$) or the cloud ($x_t^e = 0$).

\vspace{0.2cm}
\textbf{Execution Cost.}
The task's execution cost depends on the task's complexity ($F_t$) and the server's resources. Let $f^c$ and $f^e$ denote the resources at the cloud and the edge server, respectively. The execution cost is defined as shown in Eq.~(\ref{eq:computation_cost}).

\begin{equation}
	\label{eq:computation_cost}
	\chi(t) = x_t^e.\frac{F_t}{f^e} + (1-x_t^e).\frac{F_t}{f^c}
\end{equation}

\vspace{0.2cm}
\textbf{Reuse Cost.}
When the edge server employs computation reuse, we differentiate two cases:
\begin{itemize}
	\item {\em full computation reuse}: the execution cost is equal to the lookup cost to find a match in the Reuse Store Table,
	\item {\em partial computation reuse}: the execution cost is equal to the sum of lookup process and rest of computation to finish the whole task's execution.
\end{itemize}

The Reuse Store Table is a data structure that stores information related to the previously executed tasks. Each entry in the Reuse Store Table contains the name of the invoked service, the task's input data, and the task's output.
The reuse cost is then defined as shown in Eq.~(\ref{eq:cost_reuse}).

\begin{equation}
	\label{eq:cost_reuse}
	\eta(t) = r_t.L + (1-r_t).(L + \chi(t'))		
\end{equation}

\noindent where $r_t$ denotes whether a task $t$ is satisfied via a full computation reuse ($r_t = 1$) or a partial computation reuse ($r_t = 0$). $t' \subset t$ is a part of the task $t$ to be executed after performing partial reuse.

\vspace{0.2cm}
Finally, the overall task's completion cost ($\zeta_t$) is defined as shown in Eq.~(\ref{eq:cost_total}).

\begin{equation}
	\label{eq:cost_total}
	\zeta_t = \Gamma(t) + (1 - \gamma_t).\chi(t) + \gamma_t.\eta(t)
\end{equation}

\noindent where $\gamma_t$ denotes if computation reuse ($\gamma_t = 1$) or computation from scratch is applied ($\gamma_t = 0$). 

The objective function shown in Eq.~(\ref{eq:obj}) tends to minimize the overall task's completion time. The computation constraint is shown in Eq.~(\ref{eq:c1.1}), where the afforded tasks' computation at the edge should not exceed the maximum available resources. Similarly, Eq.~(\ref{eq:c1.2}) shows that the utilized bandwidth by tasks' input should not exceed the available bandwidth between users and edge. Finally, the non-negativity constraint is shown in Eq.~(\ref{eq:c1.3}).

\begin{mini!}|l|[3]
	{t, \gamma}{\zeta_t}{}{}\label{eq:obj}
	\addConstraint{\sum_{t \in T} F_t \leq f^e \qquad \forall e \in E}\label{eq:c1.1}
	\addConstraint{\sum_{t \in T} I_t \leq b^e \qquad \forall e \in E}\label{eq:c1.2}
	\addConstraint{\gamma_t \in \{0, 1\} \quad \forall t \in T}{}\label{eq:c1.3}
\end{mini!}

\subsection{Architecture Design}
\begin{figure}[!t]
	\centering
	\includegraphics[width=\linewidth]{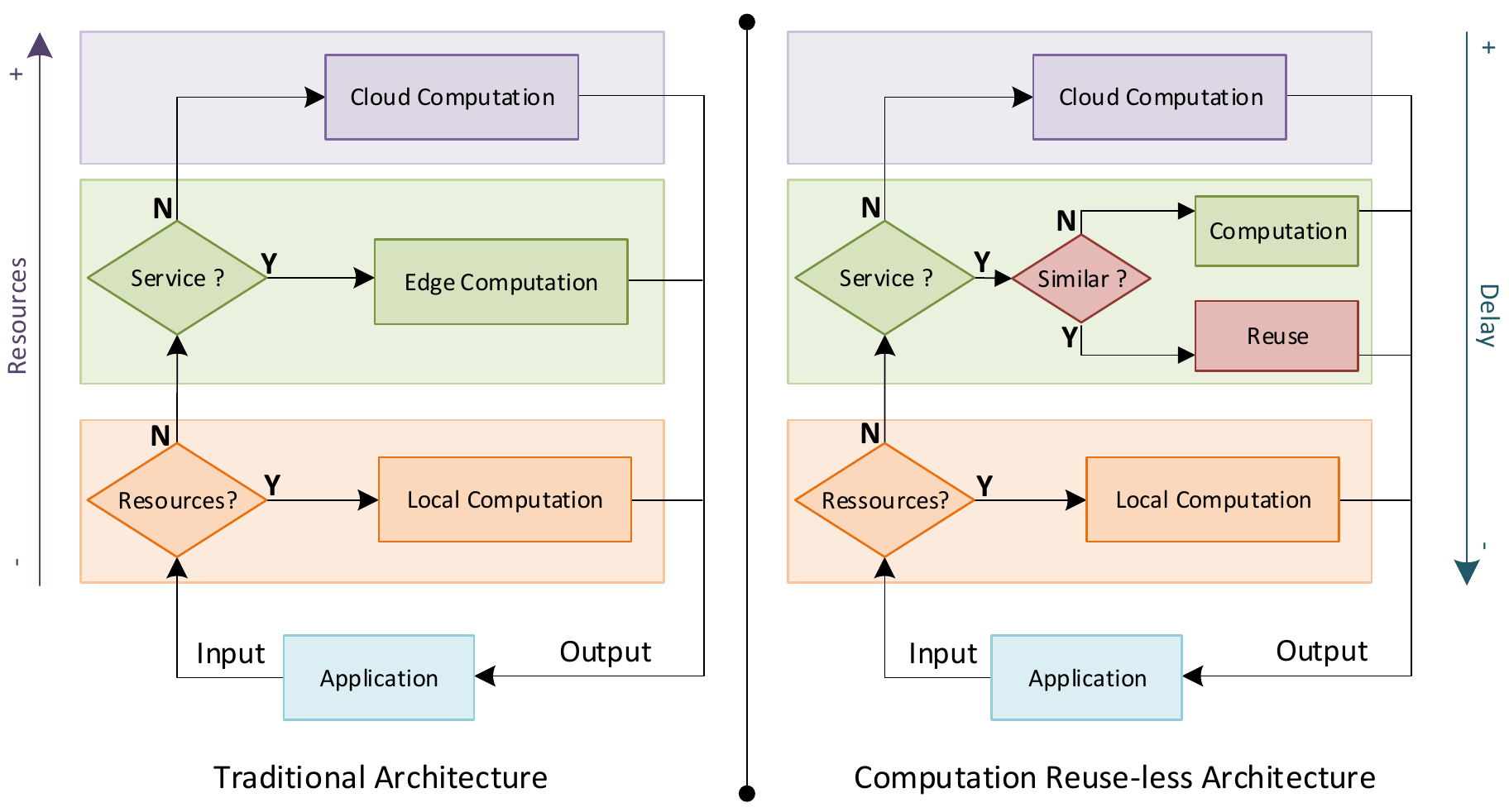}
	\caption{Computation reuse architecture at the edge.}
	\label{fig:proposed_architecture}
\end{figure}

Moving computation close to users reduces the delay, but fewer resources are available to satisfy all received tasks. Conversely, the more resources used on the distant cloud server, the greater the delay for end users.
In order to minimize task completion time while optimizing resource utilization, we have extended the edge computing paradigm with the concept of computation reuse. Unlike the traditional architecture (the leftmost part of Figure~\ref{fig:proposed_architecture}) that provides only computation at the edge, the proposed architecture (the rightmost part of Figure~\ref{fig:proposed_architecture}) adopts the computation reuse concept. The latter aims at eliminating redundant computations by storing previous outputs and later using them instead of computing from scratch every time.

\vspace{0.2cm}
\textbf{Forwarding Plane.}
Algorithm~\ref{alg:forwarding_scheme} depicts the proposed forwarding scheme at the edge. Indeed, the edge before executing a received task for an available offloaded service, checks on its Reuse Store Table if any similar task with (semi)-similar input data has already been stored. If a full match is found, the stored output data from the Reuse Store Table is used as an output for the received task and the result is sent back to the end-user. If a partial match is found, the edge uses it to satisfy part of the computation while the rest is performed via a normal computation. Otherwise (no match is found on the Reuse Store Table), a computation from scratch is performed at the edge. If the invoked service is not available on the edge, the task is forwarded immediately to the cloud for remote execution.

\begin{algorithm}[!b]
	\SetKwInput{KwData}{Input}
	\KwData{$t$: task}
	
	s $\leftarrow$ t.service()\; 
	
	\uIf{(s $\in$ offloaded services)}{
		in $\leftarrow$ t.input().feature\_extraction()\; 
		similar\_tasks $\leftarrow$ RST.query(s, in)\; 
		
		\uIf{(similar\_tasks $\neq~\emptyset$ )}{
			out $\leftarrow$ similar\_tasks.near.output()\; 
			
			\uIf{(out.partial())}{
				\textbf{return} out + s.compute(t')\;
			}\Else{
				\textbf{return} out\; 
			}
			\If{(placement(t) is True)}{
				RST.add(s, in, out)\;
			}
			
		}\Else{
			\textbf{return} s.compute(t)\; 
		}
	}\Else{
		offload(t, cloud)\; 
	}
	
	\caption{A Network-based compute reuse forwarding scheme.}
	\label{alg:forwarding_scheme}
\end{algorithm}

Yet, to design a contextual-aware computation reuse at the network level compatible with the presented forwarding scheme, different challenges need to be addressed. For instance,
\begin{enumerate*}[(i)]
	\item converting application-specific raw input data into a generic metric space,
	\item hashing multiple similar input data to the same hash codes,
	\item designing a fast nearest-match search in high dimension space, and
	\item ensuring that the output of computation reuse is accurate and correct.
\end{enumerate*}

\vspace{0.2cm}
\textbf{Similar Inputs Identification.}
The computation reuse is built mainly on the identification of similar input data that lead to the same task output. Since a task is invoking a predefined service with a determined input data, we can then identify if two tasks invoking the same service are identical, similar, or even semi-similar.

To address the aforementioned challenges, the first steps consists of extracting features from the application-specific raw input data. Various algorithms exist for feature extraction on different data such as images, video, audio, and text. 
The second step is to identify a task. Hashing functions can be used to generate a unique hash value. However, conventional hashing techniques aim to maximize  collision, which means that two similar data will have very different hash codes. This that not serve the purpose of identifying similar input. Locality-Sensitive Hashing (LSH)~\cite{andoni2008near}, on the other hand, tends to minimize the collision, which means that two similar input data can be hashed to similar hashing values. 
Calculating the distance between feature vectors in LSH allows finding similar input data. In fact, similar input data will be hashed with high probability to the same value, while dissimilar data will be hashed to different values.

\begin{figure}[!t]
	\centering
	\includegraphics[width=\linewidth]{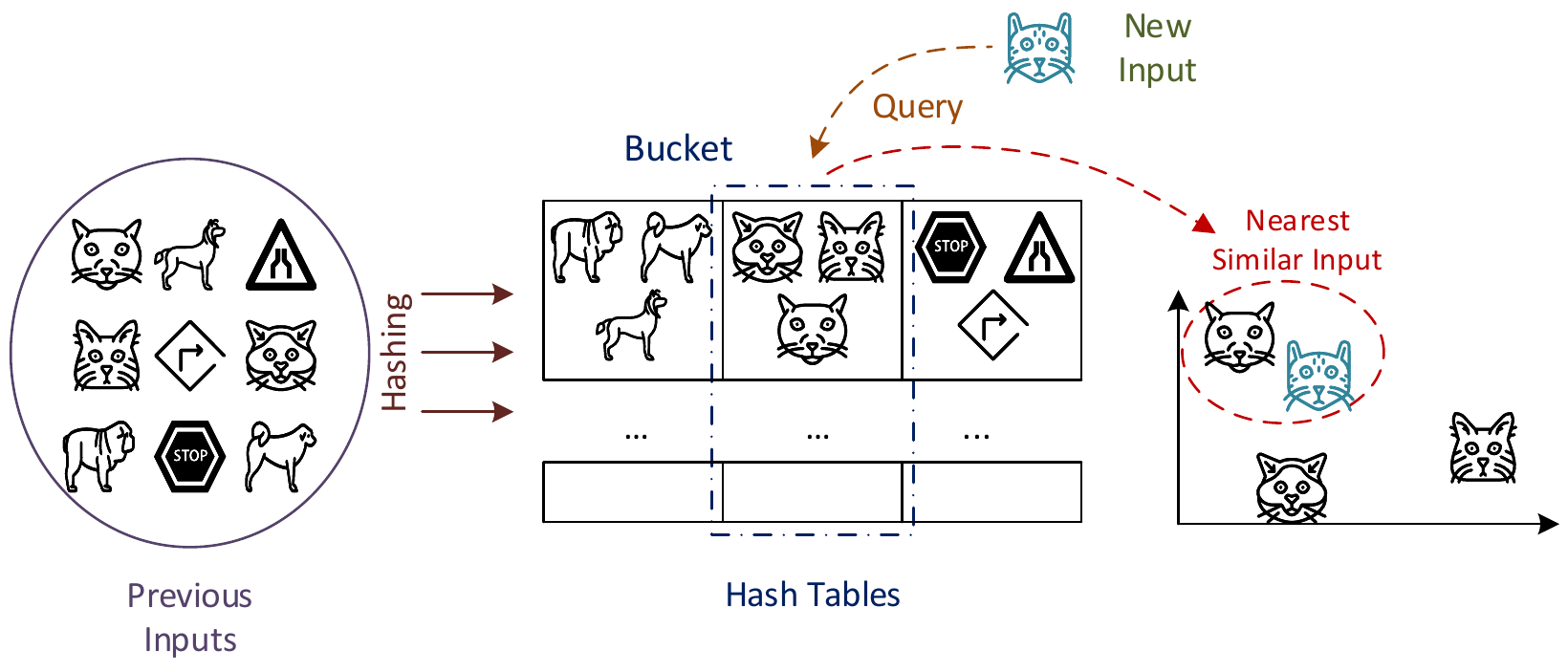}
	\caption{Input hashing and querying example in LSH.}
	\label{fig:lsh}
\end{figure}

Figure~\ref{fig:lsh} shows an input hashing and querying example in LSH.  LSH uses multiple hash tables in practice, where each hash-table employs a distinct hash function and a set of buckets (\eg dogs, cats, road signs). Similar input data will be mapped to same bucket, allowing the process to find a nearest similar input data in short time.
To find a similar input data in a hash-table, LSH locates the corresponding bucket by hashing the received input data. The entries on that bucket are candidates for similarity. The final output is selected based on a distance between the query input and the candidates in the bucket.

\vspace{0.2cm}
\textbf{Reuse Store Table.}
Hash-tables in LSH are considered as a Reuse Store Table. Indeed, each element has the hash value of the input and the output result. The table will keep adding new entries, which will affect the resource usage and the lookup process. To keep the Reuse Store Table compact and fresh, it is mandatory to remove non popular inputs (candidates) after reaching some non-reusability value. We define the Frequency metric as the number of times an entry has been reused for a period of time. Hence, we apply {\em Least Frequency Used} (LFU) policy in order to clean the table by removing the entries that have a smaller frequency rate and keep room for popular inputs.

\section{Evaluation Performance}
\label{sec:evaluation}
This work's main objective is to enhance the task's completion time and optimize resource utilization by eliminating redundant computations. It is also vital to ensure the correctness of the output when computation reuse is applied. In this section, we present the evaluation performance to gauge the effectiveness and efficiency of the proposed architecture.

\subsection{Experimental Setup}
In the experimentation, we adopted an object detection use case. The choice of this use case is motivated by the fact that it is widely used in smart cities and IoT applications, has multiple repeated invocations with similar input data, has a heavy computation routine, and requires a short response time. A similar experimental architecture can be used with other use cases and applications without any changes in its fundamental and working principles.

\begin{figure}[!b]
	\centering
	\includegraphics[width=\linewidth]{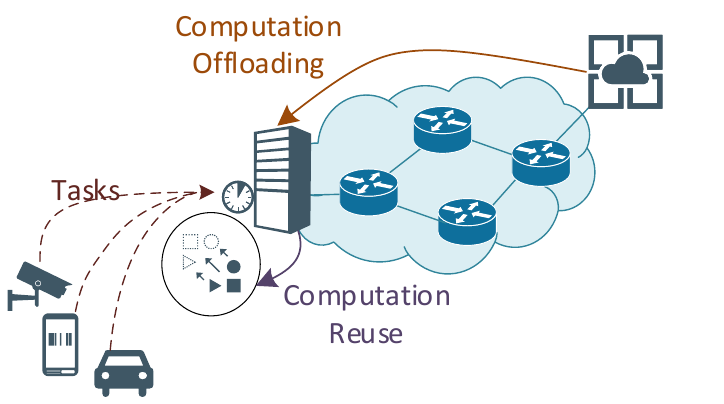}
	\caption{Evaluation network.}
	\label{fig:evaluation_network}
\end{figure}

As shown in Figure~\ref{fig:evaluation_network}, we implemented a proof of concept with three layers, \eg end-users, edge, and cloud. The service provider offloads its services from the cloud to the edge server based on the available resources at the edge. End-users send their captured images to the edge for object detection. As stated before, the edge server stores the previous computations for the possibility of reusing them in the future. We implemented LSH algorithm at the edge to hash input data, and we used LFU to maintain the Reuse Store Table. The computation can be done in three different models:

\begin{itemize}
	\item In case the edge server does not have the requested service, all tasks will be forwarded to the cloud server for remote execution. The server has unlimited computational resources yet is located far from end-users;
	
	\item Task computation will be performed at the edge server if the target service is already offloaded to the said server. However, the server has less resources for computation than the cloud server; 
	
	\item if computation reuse is applied, the edge server will use the previously stored output to satisfy the task computation. Otherwise, the computation is performed from scratch.
\end{itemize}

\begin{figure}[!b]
	\centering
	\includegraphics[width=\linewidth]{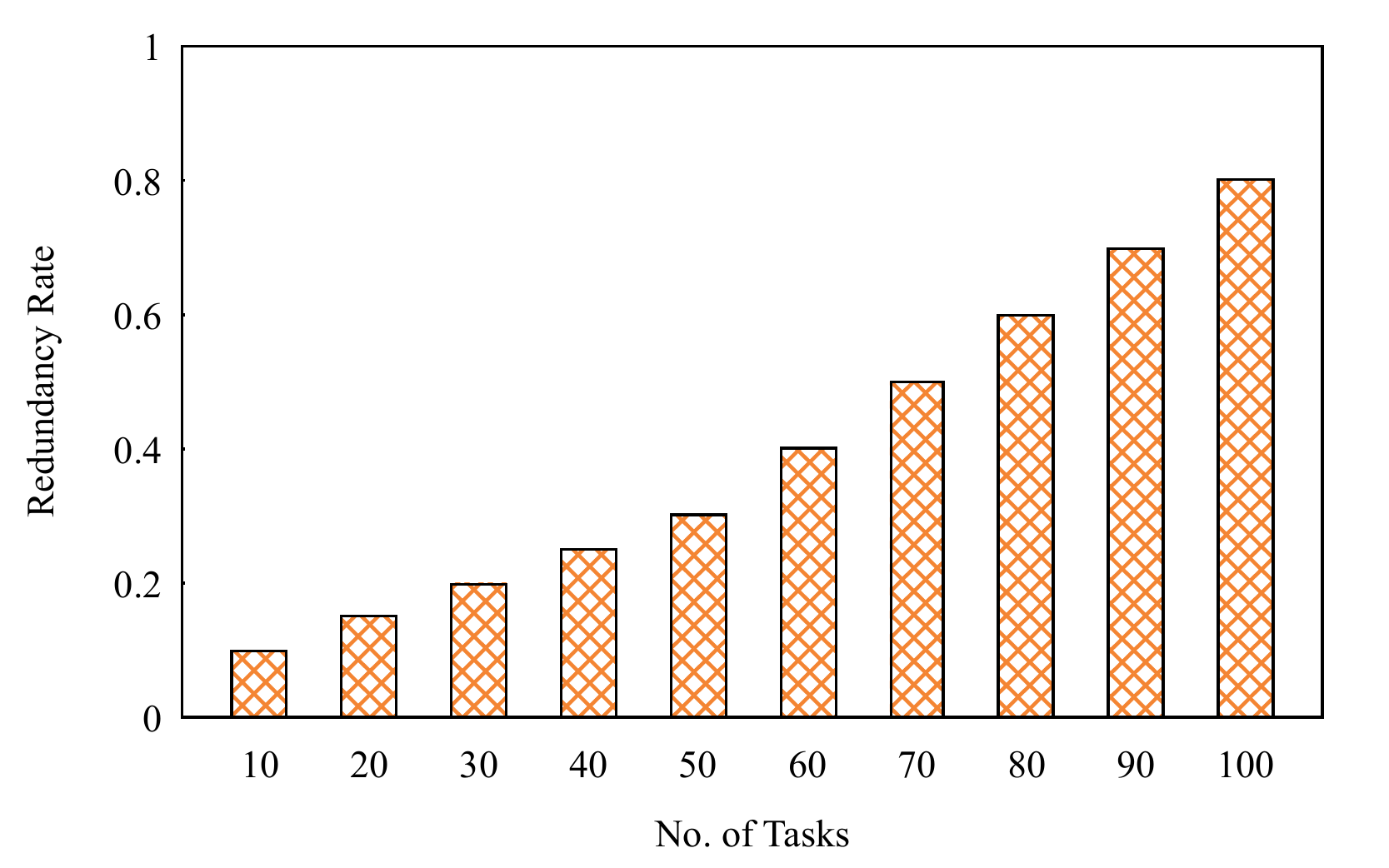}
	\caption{Redundancy rate.}
	\label{fig:redundancy_rate}
\end{figure}

We used \textit{ImageNet}~\cite{ILSVRC15}, a standard image dataset. The dataset contains over 4000 types of labeled images taken from different viewpoints and lighting conditions. We used the {\em Yolo} framework for the object detection service~\cite{redmon2016you}.
To ensure invocations with similar input, we generated a task list with a redundancy rate. Redundancy here means that the same object is repeated multiple times in the list of object detection requests. Note that redundancy refers to the same object in the given input images, not the same image, which is a realistic scenario. Figure~\ref{fig:redundancy_rate} depicts the redundancy rate for each set of requested tasks. When the number of tasks increases, the service witnesses an increase in the repeated redundant invocations. We present the \nth{90} percentile of the results collected after 10 trials.
Table~\ref{tab:parameters} provides a summary of the used parameters.

\begin{table}[!t]
	\centering
	\makegapedcells
	\caption{Experimental parameters.}
	\label{tab:parameters}
	\begin{tabular}{l l}
		\toprule
		\textit{Parameter} &
		\textit{Value} \\
		\midrule
		Hops to the edge & $1$ \\
		Hops to cloud & $[5-8]$ \\
		Edge capacity & $10-20$ tasks simultaneously \\
		Number of tasks & $[10-100]$ \\
		Tasks arrival rate & $3-10$ tasks per seconds \\
		Input redundancy & $[10-80]\%$ \\
		Replacement policy & $LFU$ \\
		Dataset & ImageNet \\
		Number of trials & $10$ \\
		\bottomrule
	\end{tabular}
\end{table}

\subsection{Experimental Results}
To prove the efficiency of the proposed architecture and gauge its performance, we measured various metrics:

\begin{itemize}
	\item \textit{Task completion time}: The time elapsed between the generation of a task by an end-user, the execution of the task (either at the edge or the cloud), and the reception of the task's results by the end-user.
	
	\item \textit{Task computation time}: The time elapsed between the reception of the task by the edge or cloud, and the production of the result.
	
	\item \textit{Task waiting time}: The time elapsed between the reception of a task by the edge/cloud and the beginning of its execution. 
	
	\item \textit{Resource utilization}: The percent of the capacity of the resources utilized to perform the computation.
	
	\item \textit{Computation load}: The amount of computation executed at the cloud, the edge, or the edge after performing reuse to satisfy the received tasks.
	
	\item \textit{Reuse gain}: The amount the gain in terms of delay and resources earned by adopting computation reuse compared to computation from scratch at the edge server.
	
	\item \textit{Computation correctness}: The percentage of achieving accurate output data (results) similar to computation from scratch. The highest the correctness, the better the accuracy.
\end{itemize}

\begin{figure}[!t]
	\centering
	\includegraphics[width=\linewidth]{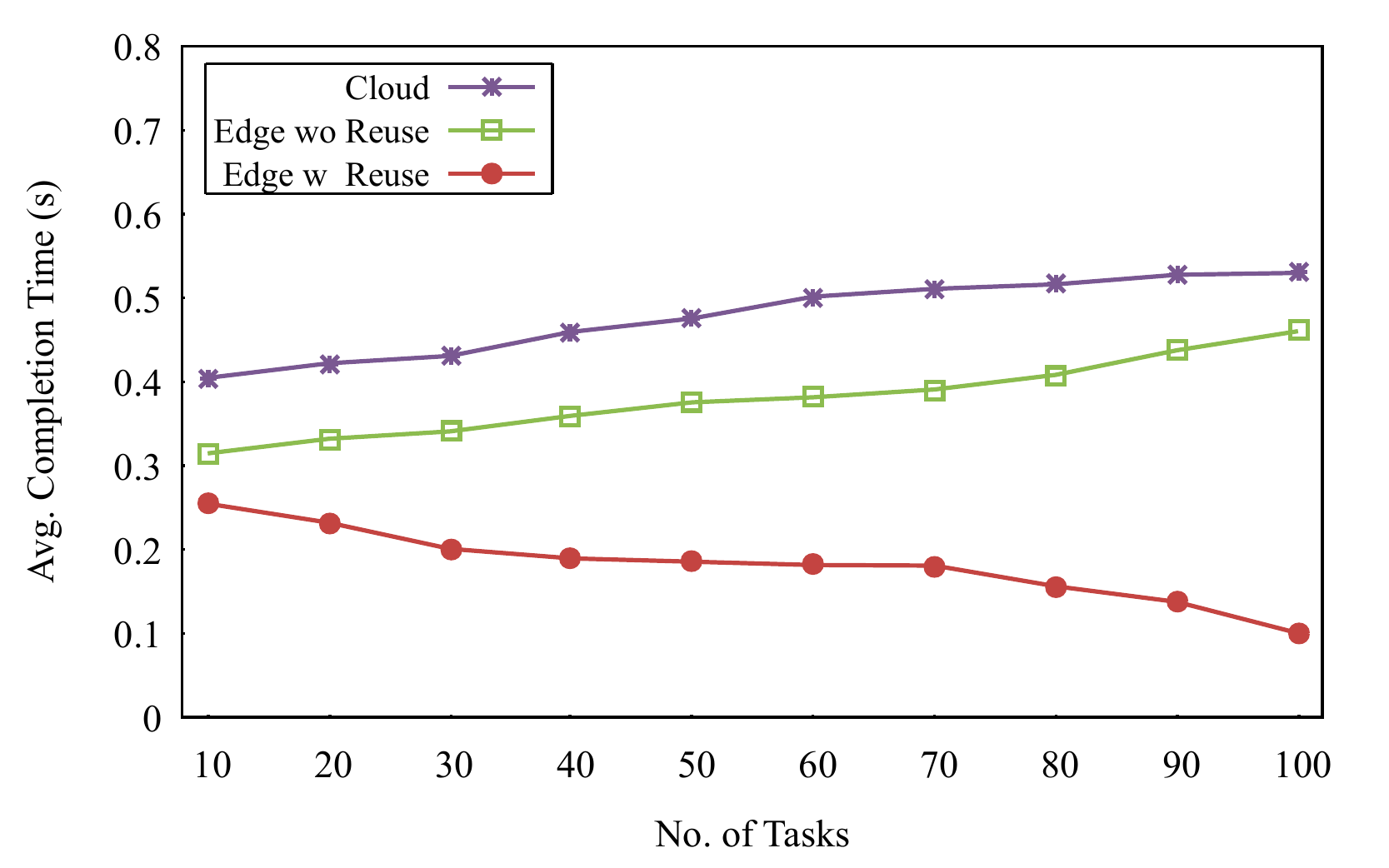}
	\caption{Completion time.}
	\label{fig:completion_time}
\end{figure}

\vspace{0.2cm}
\textbf{Task Completion Time.}
Figure~\ref{fig:completion_time} outlines the results for task completion time for both cloud, edge, and edge with computation reuse. We can notice that the cloud has the most considerable completion time because the server, where the computation is performed, is located far away from the end-users.  Thus, more time is needed to transfer the large size of input data via the core network. Conversely, the edge server is located near to end-users where less data is traversing the networks. Hence, the completion time witnesses better results compared to the cloud. Finally, computation reuse at the edge helps in avoiding redundant computation by performing only a lookup on hash-tables instead of computations, which leads to a better completion time (up to $80\%$). This is because the received tasks witness a growing redundancy rate that accelerates the task completion time.

\begin{figure}[!b]
	\centering
	\includegraphics[width=\linewidth]{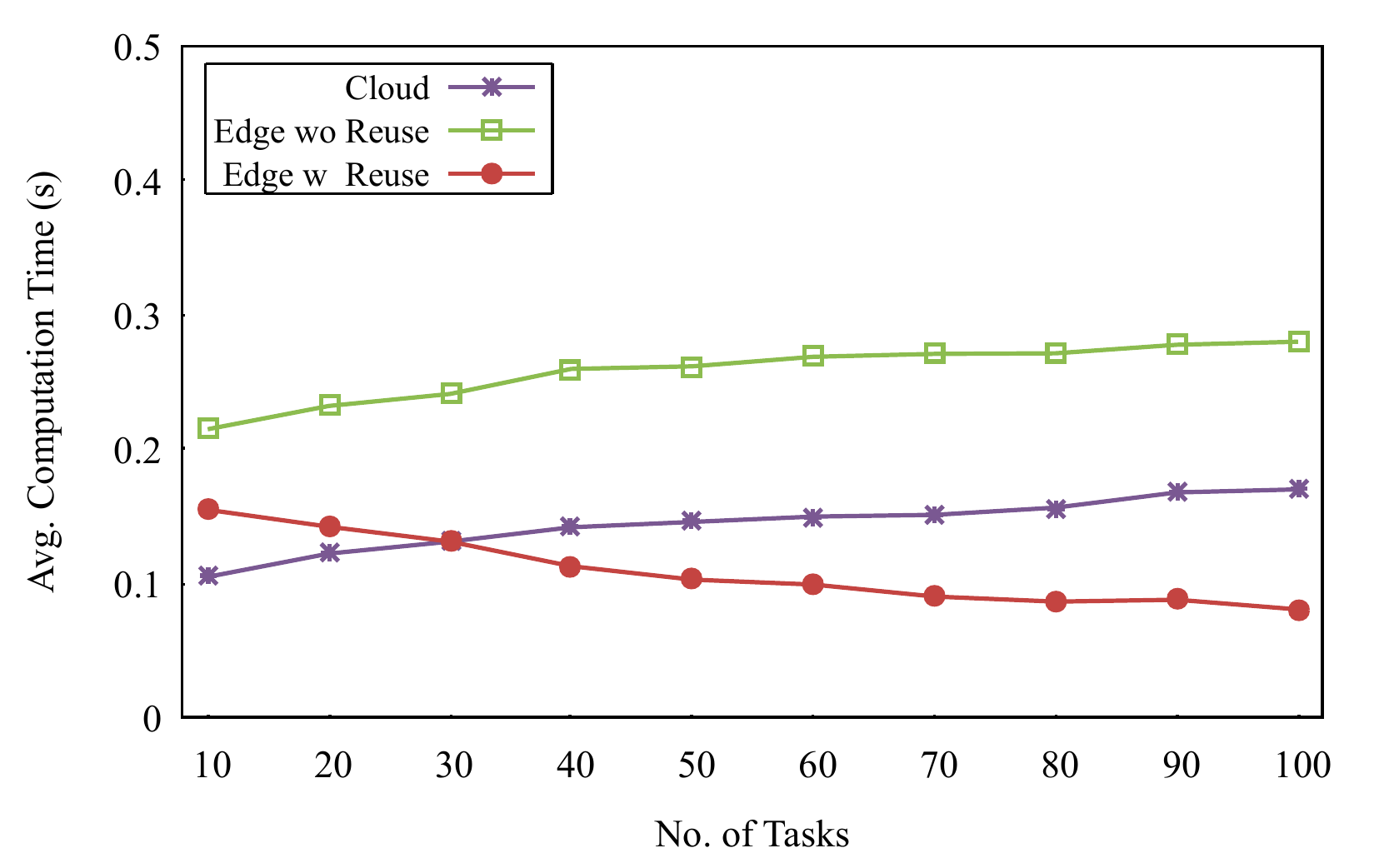}
	\caption{Computation time.}
	\label{fig:computation_time}
\end{figure}

\begin{figure}[!t]
	\centering
	\includegraphics[width=\linewidth]{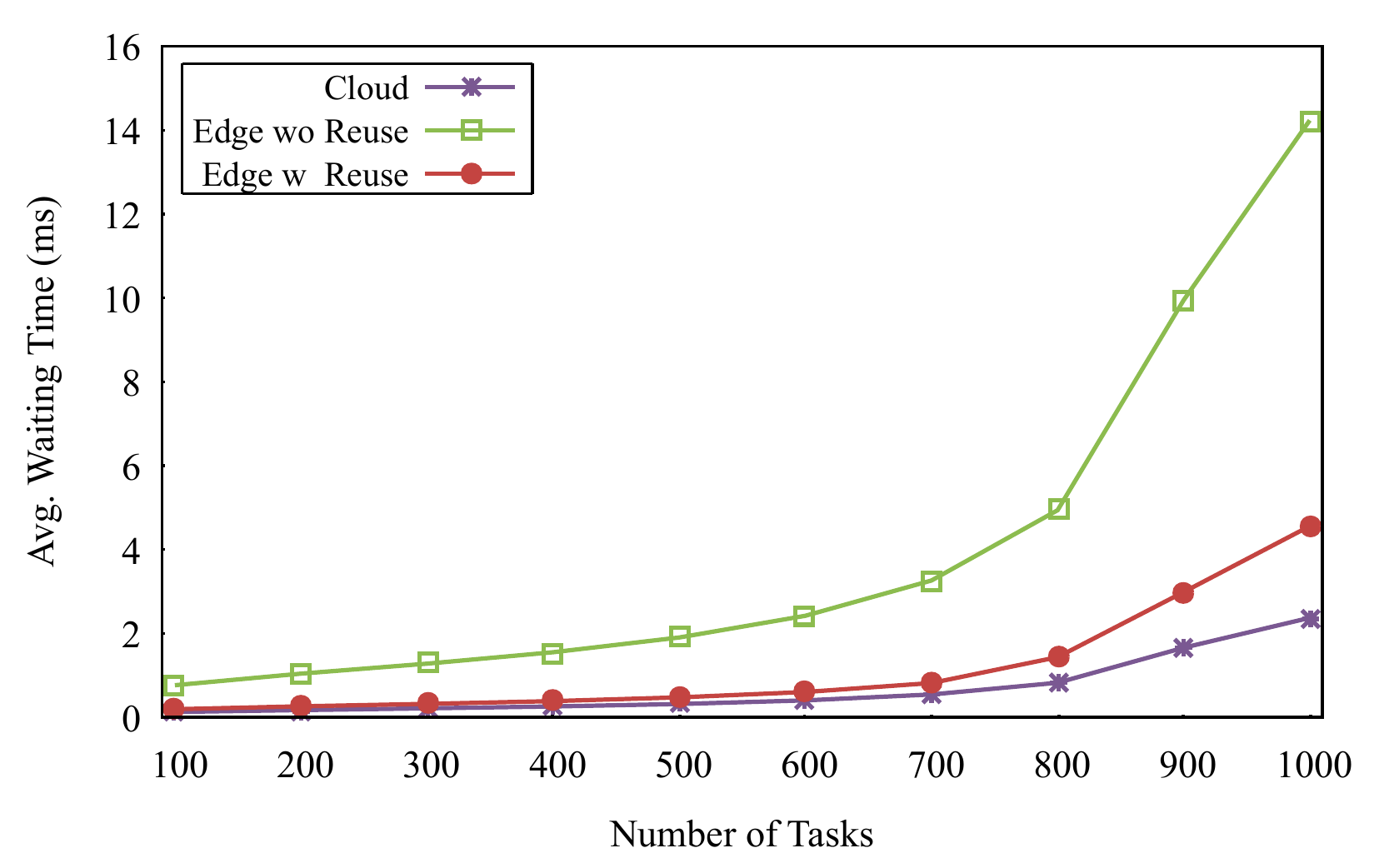}
	\caption{Waiting time.}
	\label{fig:waiting_time}
\end{figure}

\vspace{0.2cm}
\textbf{Task Computation and Waiting Time.}
Figure~\ref{fig:computation_time} shows the average computation time. The results prove that the transmission time has a high impact on cloud computing performance even with unlimited computation resources (Figure~\ref{fig:completion_time}). In contrast, the waiting time impacts the edge server capacity to perform fast computation. Figure~\ref{fig:waiting_time} shows that the waiting time increases when the number of executed tasks grows. However, the waiting time at the edge is negligible compared to the transfer time for the cloud. In Figure~\ref{fig:computation_time}, we can also notice that computation reuse at the edge exceeds even the cloud performance by processing a very negligible lookup process to find a match on the Reuse Store Table. Object detection has a high complexity that requires significant resources, while the lookup process on hash-tables is light. When the redundancy of received inputs is high, computation reuse allows fast computation since the edge finds a quick match on hash-tables, which contrasts with the higher computation time of the task on the cloud server despite its immense resources.

\begin{figure}[!b]
	\centering
	\includegraphics[width=\linewidth]{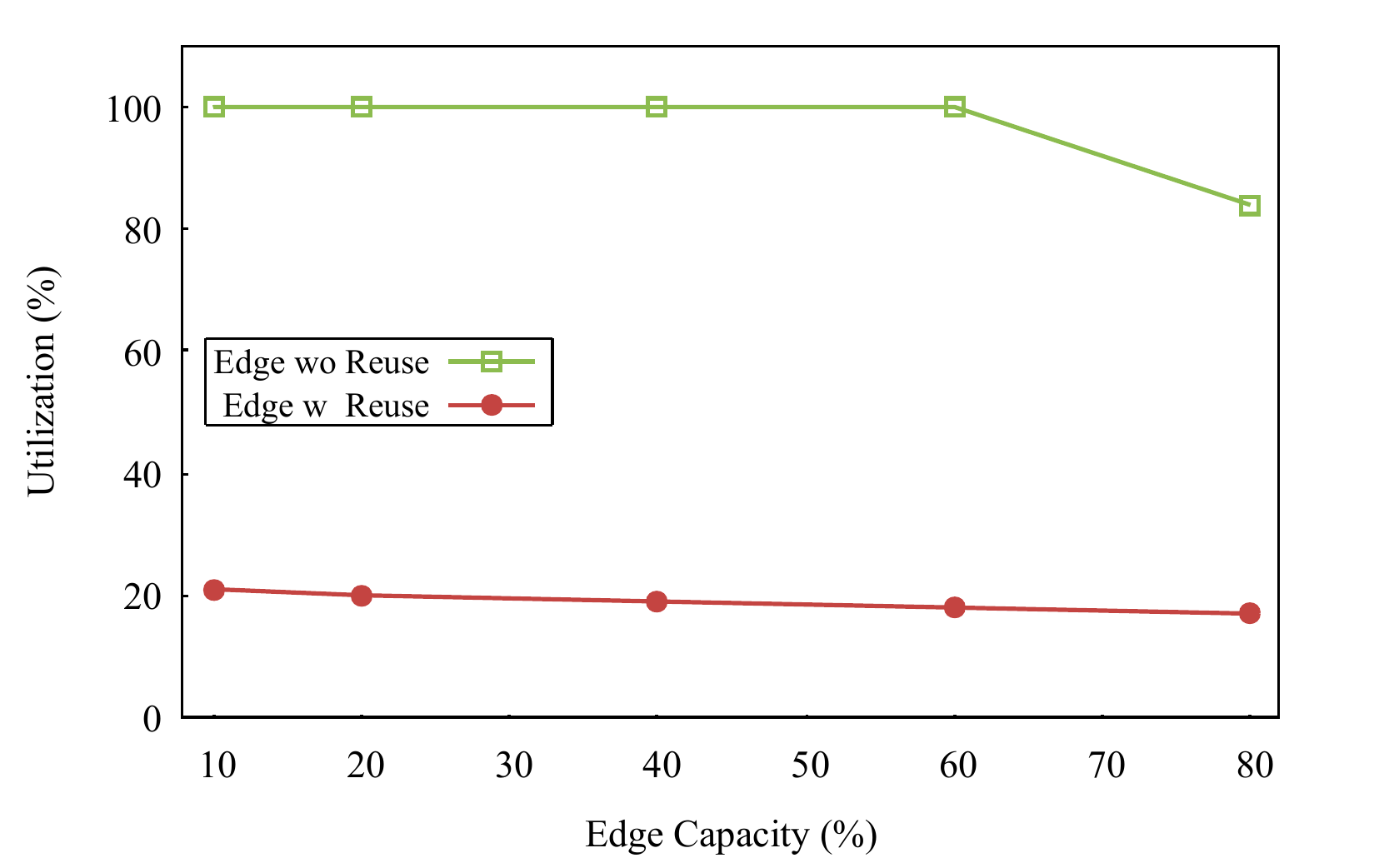}
	\caption{Resource utilization.}
	\label{fig:resource_utilization}
\end{figure}

\vspace{0.2cm}
\textbf{Resource Utilization.}
Figure~\ref{fig:resource_utilization} shows the results of computing resource utilization at the edge with and without computational reuse. In this experiment, we started by using a portion of the available edge resources, offloading services from the cloud to the edge, and measuring the resource utilization. We increased the available resources at the edge and then measured the resource usage to satisfy 1000 tasks of different services. We can notice that the edge without computation reuse uses all resources (100\%) to execute all tasks. When we increase the capacity of the server to 60\%, we can then notice a 20\% decrease in usage. On the other hand, edge with computation reuse barely reaches 20\% to satisfy all tasks since it mostly does not perform computations from scratch but only searches on hash tables.

\begin{figure}[!t]
	\centering
	\includegraphics[width=\linewidth]{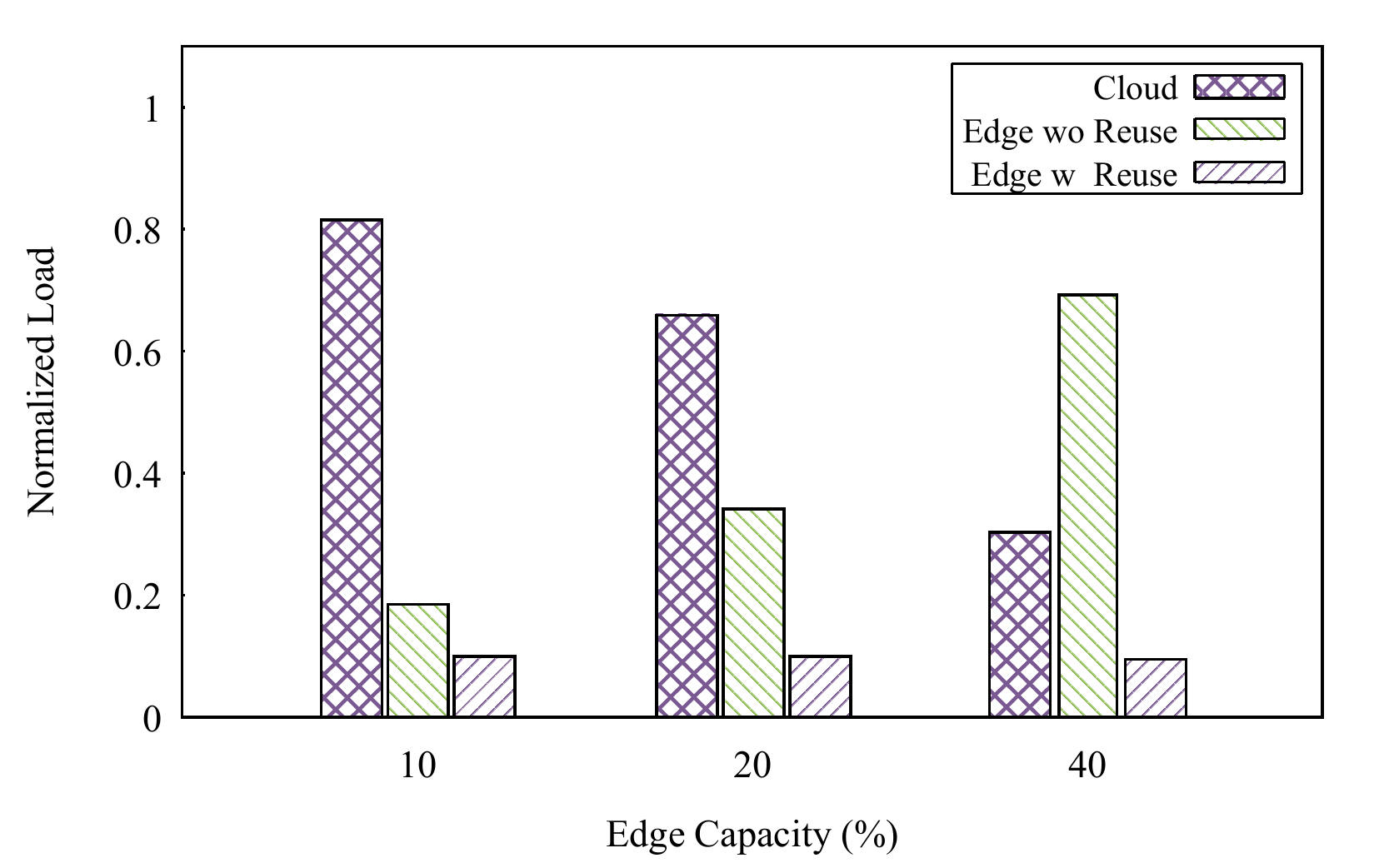}
	\caption{Computation load.}
	\label{fig:load}
\end{figure}

\vspace{0.2cm}
\textbf{Computation Load.}
In Figure~\ref{fig:load}, we present results on the load of the computing resources for 1000 received tasks. The results indicate that as we increase the capacity of the edge server, more computation is executed at the edge rather than in the cloud. Figure~\ref{fig:load} corroborates the aforementioned results (Figure~\ref{fig:resource_utilization}). For instance, with only 10\% of edge capacity, computation without reuse performs only 18\% of computation at the edge, and the rest (82\%) is executed at the cloud level. Yet, by adopting the computation reuse mechanism, the edge reuses the results of previously executed tasks, thus satisfying all tasks at the edge without bouncing the computation back to the cloud. Computation reuse reduces the overall amount of performed computation.

\begin{figure}[!b]
	\centering
	\includegraphics[width=\linewidth]{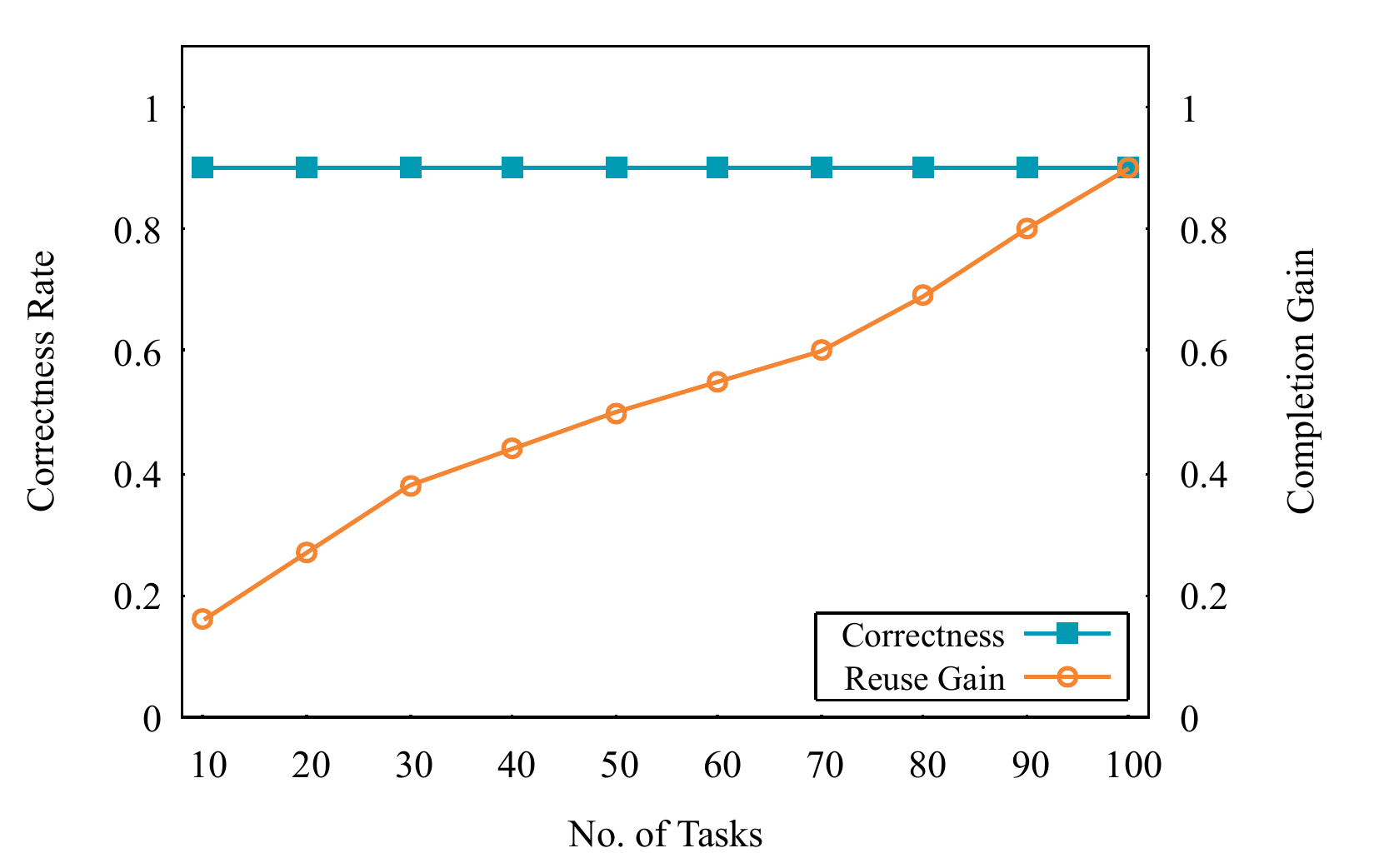}
	\caption{Reuse gain and computation correctness.}
	\label{fig:correctness_rate}
\end{figure}

\vspace{0.2cm}
\textbf{Reuse Gain \& Computation Correctness.}
Figure~\ref{fig:correctness_rate} shows that when the number of tasks increases, the redundancy increase, which in return increases the reuse gain. The more redundancy we have in the data input, less computation the edge performs, and hence few resources are utilized. The same figure shows that computation reuse is able to reach almost 0.9 of correctness rate, which means 90\% of the reuse operations were absolutely correct compared to computation from scratch. The missed 0.1 reverts to the fact that some of the computations have not been stored in the Reuse Store Table or have been evicted by the replacement policy. 

\subsection{Complexity Analysis}
The complexity of the proposed forwarding scheme is mainly based on the complexity of LSH and adding new entries to the Reuse Store Table. However, we ignore the complexity of the replacement operations (adding, updating, or evicting) since they are performed offline.

LSH maintains $l$ hash-tables, each with $n$ elements and for each element, we store a $k$-dim hash vector. Thus, the space complexity is $O(lnk)$. By ignoring lower order terms, the space complexity is $O(n^{1+\rho})$, where $\rho$ determines the time/space bounds of LSH algorithm.
For a given input $q$ to query in LSH, $O(lk)$ is the time complexity to find a match in hash-tables. For each candidate close to $q$, we spend $O(d)$ time to compute the distance $dist(p, q)$. Ignoring lower order terms, the query time is $O(n^{\rho})$.

\section{Related Work}
\label{sec:related_work}
To meet the desired quality of service (\eg delay), Edge computing allows moving services from the cloud to the edge (\ie closer to end-users) and offloading computational tasks from resource-constrained devices to Edge servers~\cite{abouaomar2021resource}. Edge technology will be transformative in many sectors of public services and industry activities. However, with the massive growth of connected IoT devices and the huge data generated, it is not feasible to move all the huge resources from the cloud to the edge to perform data analysis and processing ~\cite{duan2020convergence}.
The concept requires additional optimization techniques and enhancements to meet the continuous changes in the nature of applications and user requirements~\cite{li-rtgwg-cfn-framework-00}.  Similar to content caching, where content can be temporarily cached and used to satisfy future requests, the edge server can take advantage of computational caching.  Unlike content caching, computation caching is tied to the service logic and input data that triggers the execution routines ~\cite{irtf-coinrg-use-cases-00}.

Computation reuse paradigm~\cite{he2016exploiting, bellal2021coxnet} allows to reduce the cost of computations by reusing already executed and cached computations instead of performing the computations from scratch.  In doing so, the edge server must store the received computations (\eg service name, input data, output data), and then use them to fulfill the incoming tasks' computation~\cite{nour2021far}.

Various efforts have aimed to exploit the concept of computation caching and reuse.
The work in~\cite{drolia2017cachier}, the authors designed a system that uses the caching model to minimize latency by dividing the computation and load between the edge and the cloud.
The work in~\cite{guo2018foggycache} aims to minimize computational redundancy by using fuzzy computing. The concept focuses on caching approximate computations at the fog/edge level and then utilizes the similarities between different input to find a suitable output that fulfills the new received tasks.
Similarly, the work in~\cite{guo2018potluck} aims to store and share the processing results between applications and leverage a set of algorithms to evaluate the input similarity to maximize deduplication opportunities.
These works, in contrast to our work, are built on top of approximate computations. The approximate aspect may not lead to intact results, and thus the output accuracy is questionable. Guaranteeing computation accuracy in computation reuse is a critical aspect, especially in those applications that require rigorous results (e.g., face recognition).
The work in~\cite{mastorakis2020ICedge} uses  named data networking to extend edge computing to perform in-network computing using the name of the content and reusing the result to satisfy other similar tasks.

Other approaches such as ApproxEye~\cite{he2017approxeye}, DeepMon~\cite{huynh2017deepmon}, CBinfer~\cite{cavigelli2017cbinfer}, and DeepCache~\cite{xu2018deepcache} attempted to use the same concept to enhance the applications' performance.
ApproxEye is a framework for partial reuse of approximate computations to improve computer vision of micro-robots. The framework uses computation locality to reuse similar prior computations to reduce redundant computations. Other solutions, such as DeepMon and CBinfer, use the concept of computation reuse to reuse intermediate results from the previous image to compute the current image in a convolutional neural network for real-time video processing.

In all of the solutions mentioned above, the required correctness of the computation is subject to the service/application specification and its peculiarities. For instance, a service that allows identifying obstacles for smart vehicles requires a high accuracy of the computed results. Indeed, low accuracy results may lead to traffic accidents and put human lives at risk. Thus, some of the works presented above are not best suited to ensure such services. Conversely, for an  instant text translation service, the end-user is able to grasp the meaning of the translated text even if it is not accurately translated.

\section{Conclusion \& Future Work}
\label{sec:conclusion}
In this paper, we studied the use of computation reuse concept in IoT applications, where the same service is often invoked multiple times by multiple devices using similar input data. By reusing the results of similar already executed tasks, we can achieve lower response time and efficiently utilize the edge resources. In doing so, we extended the edge server with an extra module to perform the computation reuse. We designed a network-based compute reuse architecture. We applied a Locality-sensitive hashing algorithm to hash similar input to similar hash. Similar outputs are regrouped in the same bucket that helps to accelerate the match lookup process. The evaluation showed that by using network-based computation reuse, we were able to not only decrease the completion time up to 80\%, but also optimize resource utilization up to 60\%, while ensuring a high computation gain with high accuracy and correctness.
The diversity in the accuracy of the required results across services provides the opportunity to design service-dependent computation reuse rather than a generic-purpose computation reuse architecture. Thus, services can start independently benefiting from the computation reuse feature after validating that the current computation correctness satisfies their requirements.
In future work, we aim to design the computation reuse forwarding plane as a service, so it can be invoked not only by multiple users using the same application but across applications. We also aim to join computation offloading and reuse at the edge server, in order to offload services with high reuse gain. Finally, as data privacy is concerned, we will also investigate the privacy issue when it comes to computation sharing.

\section*{Acknowledgments}
The authors would like to thank the Natural Sciences and Engineering Research Council of Canada (NSERC) for the financial support of this research.

\bibliographystyle{IEEEtran}

\end{document}